To quote this work:

V. Brien, A. Dauscher, P. Weisbecker, J. Ghanbaja and F. Machizaud
Transversal growth microstructures of quasicrystalline Ti-Zr-Ni films
Journal of Crystal Growth, 256 (3-4), 407-415 (2003)
hal-02882479 , doi:10.1016/S0022-0248(03)01350-2

Thank you.

________________________________________________________________

# Transversal growth microstructures of quasicrystalline Ti-Zr-Ni films

## V. Brien[1], A. Dauscher[2], P. Weisbecker[1], J. Ghanbaja[3], F. Machizaud[1]

1 Laboratoire de Science et Génie des Matériaux et de Métallurgie, UMR 7584, CNRS-INPL-UHP, Parc de Saurupt, ENSMN, 54042 NANCY Cedex, FRANCE
2 Laboratoire de Physique des Matériaux, UMR 7556, CNRS-INPL-UHP, Parc de Saurupt, ENSMN, 54042 NANCY Cedex, France
3 Service Commun de Microscopie Electronique à Transmission, Faculté des Sciences - Université Henri Poincaré - Nancy I, BP 239, 54506 Vandoeuvre lés Nancy - FRANCE
For correspondence: V. Brien, E-mail: brien@mines.u-nancy.fr, Tel: 00.33.3.83.58.40.78, Fax: 00.33.3.83.57.63.00

### ABSTRACT
**Pulsed laser deposition from a Nd:YAG laser was employed in production of hundreds of nanometer thick quasicrystalline Ti-Zr-Ni films on glass substrate. The influence of deposition temperature $T_s$ on the structure, morphology and microstructure of the films across their thickness was investigated. The morphology and microstructure features were evaluated by X-ray diffraction and transmission electron microscopy techniques. The low deposition temperatures were found to produce films with nanometer sized grains embedded in an amorphous matrix. The grains exhibit quasicrystalline order. The higher deposition temperatures lead to films whose structure is not uniform all along the growth direction. The layer in contact with the substrate is a very thin amorphous layer. The main part of the film consists of crystallized columns. The columns have grown from a nano-crystallized layer where the size of crystallites increases with increasing thickness.**

## 1. INTRODUCTION

Since the discovery of quasicrystals in 1984, research in the field developed extremely quickly and intensely. Quasicrystals were found to exhibit a very specific combination of interesting and unexpected physical and chemical properties [1-11]. They are unfortunately fragile at room temperature and thus it is difficult to envisage them as massive material for a practical use. Industrial applications imply then that the material should be used, either in the shape of films, or combined with other materials such as in composites. In fact, such a technological constraint corresponds well to the constraints implied by an eventual energetic application like hydrogen storage. Indeed, the optimisation of solid batteries necessitates the optimisation of the surface/volume ratio of the constitutive material of the batteries. This fact gives hope that a possible industrial application may be found for the thermodynamically stable Ti-Zr-Ni icosahedral quasicrystalline phase that has been shown to store hydrogen in a reversible way [12-14]. It makes this phase and its structurally related compounds a possible alternative as hydrogen storage material. The excellent performance demonstrated in [12-14] and the prospect of being able to produce "clean" energetic devices justify such a study. Preliminary work has been carried out by Brien et al. on quasicrystalline Ti-Zr-Ni film growth [15]. This study shows it is possible to obtain a thick (the term thick will be used here for thicknesses of around a few hundreds of nm) pure quasicrystalline film. Understanding the stages of quasicrystalline growth is critical for the optimization and reproduction of film fabrication for further application. So, different substrate temperatures between 25 and 275 °C were applied during film synthesis. Evolution of the structure versus deposition temperature was then subsequently studied and is reported in this paper. Four-circles X-ray diffraction was systematically performed to obtain the global nature and structure of the films versus deposition temperature. Detailed microscopy investigation on cross-sections of samples deposited at 100 and 275 °C was carried out by transmission electron microscopy (TEM) in order to characterize their transversal microstructures. Discussion then leads us to put forward the growth behaviour of thick samples versus deposition temperature. One has paid particular attention to the initial growth stages. This work provides useful information on growth mechanisms of quasicrystals.

## 2. EXPERIMENTAL PROCEDURE

**Synthesis**

Films were prepared on glass substrate under high vacuum in a stainless steel chamber. A pulsed Nd:YAG (Qantel, YG 571C) laser was used for the ablation process, working at the fundamental wavelength (1064 nm), a repetition rate of 10 Hz and pulse duration of 10 ns. The laser fluence used was 72 J/cm$^2$. Experiments were performed at an initial pressure of $10^{-7}$-$10^{-8}$ mbar. The targets used for the pulsed laser deposition were obtained by cutting a $Ti_{45}Zr_{38}Ni_{17}$ ingot and were installed on a rotating holder. The ingots were prepared by melting high purity metals in an induction furnace under helium atmosphere (Radio Frequency melting). Their chemical homogeneity and dispersion of phases were determined by electron probe microanalysis (EPMA). The desired substrate temperatures ($T_s$ = 25, 100, 140, 160, 240 and 275 °C) were reached thanks to a small resistance located behind the substrate holder, and checked by a thermocouple. Typical deposition rates obtained were 1-8 Å/s. More details on the set-up of the experiment can be found in [15]. The films obtained were all strongly adherent to their substrate and their thickness were in the 0.5-1 µm range.

**Techniques of characterization**

Compositions of films were checked by EPMA and found to be equal to $Ti_{41.5}Zr_{41.5}Ni_{17}$ (atomic %, ± 1 at. %).
The films were characterized by X-ray diffraction (Co$K_\alpha$ radiation, λ = 0.178897 nm). In case a texture was present in the film, we used a four-circles diffractometer. 22 scans were thus recorded at different $\chi$ and added in order to accumulate and vizualize all the possible Bragg reflections on a same pattern. These results were obtained with $\chi$ ranging from 0 to 55° assuring a $\varphi$ rotation of the sample at 600 rotations per minute. In this way it is possible to visualize all the possible Bragg diffraction peaks in the range 30° ≤ $2\theta$ ≤ 96°.
TEM observations were carried out on a PHILIPS CM200 microscope operating at an accelerating voltage of 200 kV. Cross-sections of the samples deposited at $T_s$ =100 °C and 275 °C were prepared by ion milling with GATAN PIPS equipment samples that were mechanically pre-thinned and embedded.

## 3. RESULTS
### 3.1 Microstructure of the thick films versus deposition substrate temperature $T_s$
#### 3.1.1 Four-circles X-ray diffraction

Four-circles X-ray diffractograms were obtained from the films respectively deposited at $T_s$ = 25, 100, 140, 160, 240 and 275 °C and are given in Fig. 1. The indexation of the peaks was placed on the figures. The system of indexation created by Cahn et al. [16] was used for the quasicrystalline phase.
At low temperatures (25 and 100 °C), the patterns show signals exhibiting broad maxima at around 43 and 78 ° characteristic of an amorphous and/or nano-crystallised phase (Fig. 1(a) and (b)). The first and most intense maximum is in both cases localized around the positions of the main peaks ([2/3 0/0 1/2] and [2/4 0/0 0/0]) of the icosahedral Ti-Zr-Ni structure. As TEM investigation will show (cf. § 3.1.2), the films prepared at 25 and 100 °C are in fact made of nanocrystallites embedded in an amorphous matrix (called domain I). Profile fitting of the broad peak located at 43 ° with the two main peaks [2/3 0/0 1/2] and [2/4 0/0 0/0] of the icosahedral structure was then attempted for both temperatures. The best fit is obtained by taking into account a third broad maximum which confirms the presence of an amorphous phase. The results are given in Fig. 2 and table 1. Using the Scherrer formula based on the value of the full width at half maximum (FWHM) of the respective peaks, the evaluation of the corresponding nano-crystallites'sizes was obtained and is given in table 1. Calculation was carried out assuming that no micro-strain, or lattice distortion widen the peaks. The average particles size is 2 (±1) nm for the film prepared at 25 °C and 3 (±1) nm for the one prepared at 100 °C.

Above 100 °C, one can see the appearance of a set of fine and intense reflections that can be attributed to the primitive (P-type) hypercubic lattice of the icosahedral phase (icosahedral group $m\bar{3}\bar{5}$, domain II). The FWHM of the peaks remains approximately the same whatever the $T_s$ is. A slight evolution can however be noticed with increasing $T_s$ (Fig. 1(c) to Fig.1(f)).

The less intense peaks of the icosahedral structure which are quasi-absent at 140 °C appear at high temperatures (typically the set of peaks at 70 ° and just above 80 °), or are present at low temperatures but get more resolved at higher $T_s$ (i.e. peaks located just above 90 °). This indicates that the coherence length of the quasicrystalline order increases as Ts increases. The lattice parameters ($a_{6D}$ and $a_q$ of the icosahedral structure) have been calculated against $T_s$ and are given in table 2, showing they increase with $T_s$. On top of all this, one can see the nano-quasicrystalline phase which grows at low $T_s$ is still present above 100 °C as the two broad maxima of diffusion background are still visible at 2 θ = 43 and 78 °. Its presence diminishes with increasing $T_s$ and its related diffusion backgrounds completely disappear at 275 °C.

In order to characterize the respective morphologies and prove the domain (I) is nano-crystallized as quoted above, the transversal microstructure of the films prepared at 100 and 275 °C were studied by TEM

**3.1.2 TEM study of the cross-sections prepared from the films deposited at 100 and 275 °C**
TEM imaging is a direct method of observation and gives a clear idea of the transversal structure of the samples prepared at 100 and 275 °C.

Fig. 3 shows the recorded image and diffraction pattern on the cross-section sample prepared from the film deposited at 100 °C. The electron diffraction pattern (Fig. 3(b)) reveals two diffuse halos, characteristic of an amorphous phase. A widened and intense ring of diffusion is superimposed on the first halo and may be related to a nano-crystallized phase. The presence of nanocrystallites embedded in an amorphous matrix is confirmed on the dark field image (Fig. 3(a)). The distribution and the size of these nanocrystallites (1 to 2 nm) are homogeneous in the whole thickness of the film. TEM observations on the sample deposited at $T_s$ = 25 °C confirm this observation. The size of the crystallites is however slightly smaller (cf. table 1).

An overall view of the cross-section sample prepared from the film deposited at 275 °C is shown in Fig. 4(a) (dark field image) and 4.b (bright field image). The first stages of growth (Fig. 4.(b)) consist of two layers called A and N. The first 5 nm thick layer (A) in contact with the substrate appears to be without contrast, characteristic of an amorphous layer. The second layer (N), which is 60 nm thick, is more contrasted and clearly made of particles (Fig. 4 (a) and (b)). We noted that the size of the nano-grains progressively increases the further they are from the substrate. The biggest size of grains found is ≈ 6 ± 1 nm. A quite dense structure follows: the bulk of the film (80 % of the total volume) is made of columns that are perpendicular to the normal direction of the films and stick well together. The average width of the columns is around 50 nm. Two electron diffraction patterns were recorded (Selected Area Diffraction) using two sizes of apertures (Fig. 4(c) and (d)). The two selected areas were sketched in Fig. 4(a) (dotted circles). The diffraction pattern that corresponds to the biggest area is shown in Fig. 4(d): all the rings have been indexed according to the icosahedral structure. In the second diffraction pattern (Fig. 4(c)) we notice lines of τ related spots (τ gold number, characteristic of quasicrystalline structures). These quasicrystalline columns correspond to the icosahedral domain identified at high temperatures by X-ray diffraction in § 2.1.1.

There were two main stages of the film prepared at 275 °C. The first stage corresponds to the set-up of a domain like domain (I) amorphous matrix containing nano-grains whose size progressively increases with thickness. The second corresponds to the set-up of quasicrystalline columns. The Fig. 4.(e), which is a cut going through the median plane of the column, clearly shows the growth of the column starts from a germ of 6 nm (labelled G) and is initially cone shaped. As TEM imaging of the cross-section shows, the ratio of volume of domain (I) in the film over the volume of domain (II) is very small. It explains why the signature of domain (I) does not appear on X-ray diagrams at 275 °C (Fig. 1(f)).

So, two types of growth morphology are found. $T_s$ = 100 °C leads to a homogeneous nano-crystallized film, while $T_s$ = 275 °C leads to a more complex morphology.

**4. DISCUSSION**
A fair idea of the growth microstructure of the films for any used $T_s$ can be deduced crossing observation of X-ray results with those of TEM results.

X-ray diagrams show the structure of the films is progressively modified with increasing $T_s$ (Fig. 1):

At low $T_s < 140$ °C, the films are composed of a nanocrystallized phase embedded in an amorphous phase (domain I). The nanocrystallites'size increases and the amorphous phase disappears progressively when $T_s$ increases.

At high $T_s \geq 140$ °C, the domain (I) (amorphous phase + nanocrystallites) is in the films at 140 °C and still present by TEM but no longer detectable by X-ray diffraction at 275 °C. An icosahedral phase (domain II) is increasingly important with increasing $T_s$.

TEM gives a morphological description and confirms the existence of domains (I) and (II) which are related to two growth modes respectively called (I) and (II) (Fig. 3 and 4):

At low $T_s$, the domain (I) is homogeneous and sets up in the whole thickness of the films, while the nanocrystallites are of regular size and uniformly dispersed in the amorphous matrix (Fig. 3).

At high $T_s$ the domain (I) of thickness t is always the first to grow from the substrate (except the first layers that are amorphous, Fig. 4(b)). To match the X-ray results at high temperatures, t must decrease with increasing $T_s$. The columnar domain (II) is clearly growing from domain (I) and from the grains which exhibit the biggest size (**6** nm). One can conclude the switch from growth mode (I) to growth mode (II) occurs only if $T_s$ is high enough (Ts > 100 °C). The switch is determined by the possibility of columnar germination. This happens when the size of nanograins of domain (I) has reached a threshold value. Results show this threshold value is reached as quick as $T_s$ is high. The measurements of the nano-grains show the last layers before columnar growth at 275°C are made of grains whose size is around 6 nm. This value has to be compared with the size of atomic clusters in the icosahedral structure. These kind of structures are composed of two major types (Bergman, Mackay) depending on the icosahedral structure [17-21]. The work done by X-ray data Rietveld refinement of the W-approximant of the Ti-Zr-Ni quasicrystalline phase show it is of Bergman type [22]. This cluster is also the base cluster of the icosahedral structure, it is made of 44 atoms and its radius -from the centre of the cluster to the centre of an atom of the last shell- measures $a_q$ ($a_q = 0.531$ nm, table 2). The structure is a hierarchical arrangement of clusters, and every hierarchical level is obtained by inflation of the one before. Consequently, the diameter of the cluster related to the hierarchical level n is given by $d_n = 2a_q \tau^{3n} + d_{n-1}$, n $\in \mathbb{N}$, n $\geq 1$, $d_0 = 2a_q + d^*$, $d^* = 0.296$ nm taken as the average radius of atoms in the structure. One obtains $d_0 = 1.062 + 0.296 = 1.362$ nm, $d_1 = 4.498 + 1.362 = 5.860 \approx 6$ nm. It seems, then that the critical size to be overcome before growth switches to columnar growing, is of the order of magnitude of a cluster of the first inflation of the base Ti-Zr-Ni cluster.

To sum up, low temperatures lead to homogeneous nano-quasicrystalline films. Higher temperatures bring more energy and allow above a certain thickness t the set-up of a more ordered structure: 50 nm columns of quasicrystalline structure whose coherence length gets bigger with increasing $T_s$. The thickness of domain (I) t is as small as $T_s$ is high.

So, the applied substrate temperature during the growth of films by laser ablation allows the direct control of the microstructural quality of films: on one hand a homogeneous nano-quasicrystalline microstructure, on the other hand a microstructure made of quasicrystalline columns with an underlying nano-quasicrystalline part whose thickness can be chosen by varying $T_s$.

The films were found to exhibit an excellent adherence to the glass substrate. This adherence can be explained by the presence of the domain (I) and the amorphous 5 nm thick layer which is an the interface between the columnar domain (II) (main part of the film) and the glass substrate. Indeed, the amorphous phases are generally able to relax some mechanical stresses thanks to their free volume. This mechanical relaxation is further assisted by the possible relaxation of the multiple grain boundaries from the nano-crystallized zones.

These columns stick well together and typically look like columnar morphologies obtainable by growing metallic films through evaporation at high temperatures (see the structural descriptions in the model established by Movchan and Demchishin, and Thornton [23-24]).

## 5. CONCLUSIONS

Pulsed laser deposition is therefore an efficient process for growing well adherent good quasicrystalline samples.

Two techniques of investigation were necessary to fully characterize the samples and confrontation of both type of results (TEM and X-ray diffraction) successfully gave a fair idea of the structure, morphology and microstructure of the films across their thickness versus the substrate temperature.

The substrate temperature was revealed to be a critical parameter determining the structure and microstructure of the films.

The first stages of growth systematically adopt a nano-crystalline morphology that grows on an extremely thin amorphous layer. The increase of the size of the crystallites with thickness during the first stages condition the switch to a second columnar growth mode and occurs more easily the higher the substrate temperature $T_s$. The switch to the second growth mode is quicker, the higher the $T_s$. No increase of crystallites'size is observed with increasing thickness at low $T_s$ and films are homogeneously constituted of nano-crystallites whose size also depends on $T_s$, being smaller at small $T_s$.

**Acknowledgments:**

The authors would like to thank Dr R. Ralani from GATAN for preparing the cross-section samples.

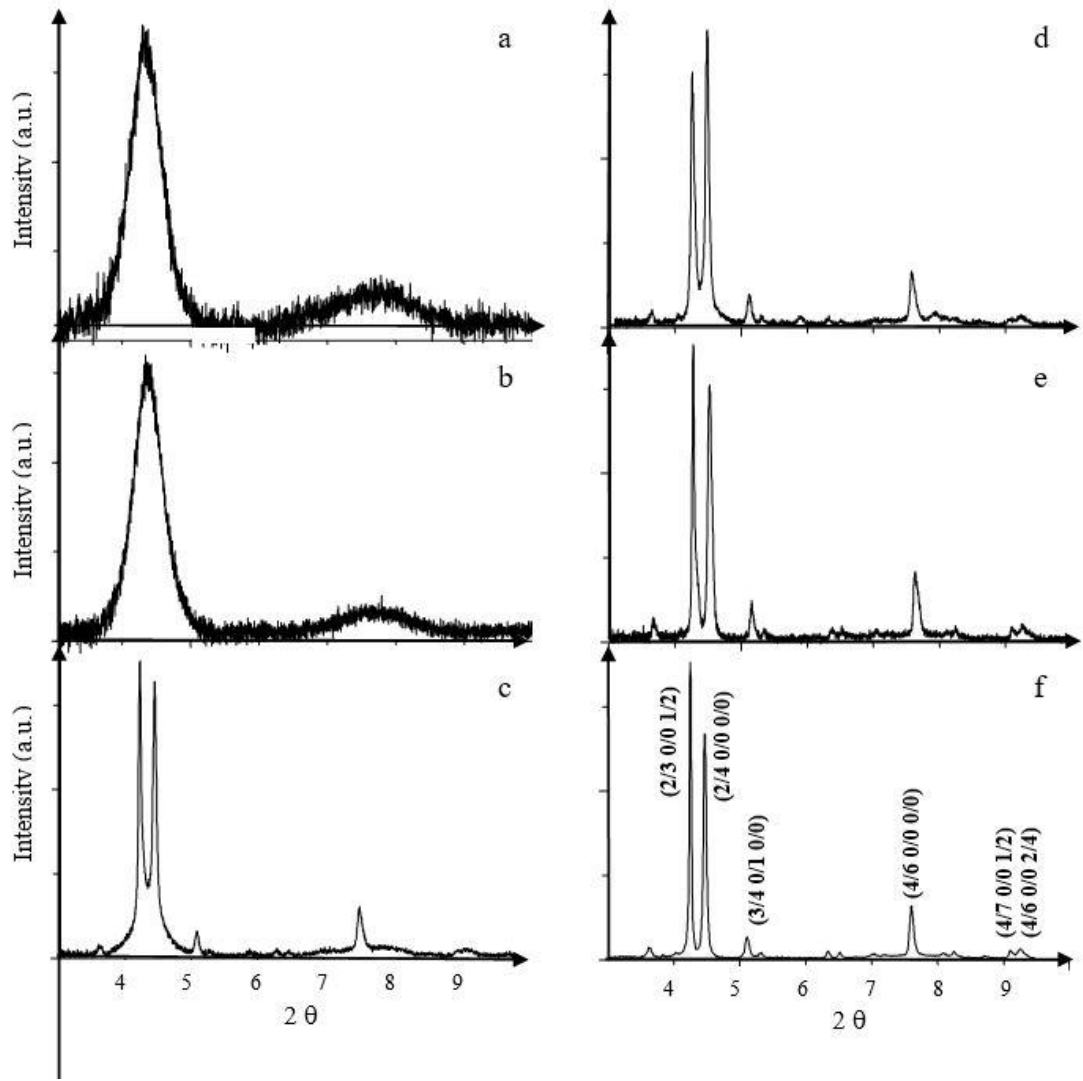

Figure 1: X-ray diffraction patterns obtained on Ti-Zr-Ni thick films deposited at a/ 25, b/ 100, c/ 140, d/ 160, e/ 240 and f/ 275 °C (Four-circles X-ray diffraction). a.u. stands for arbitrary units.

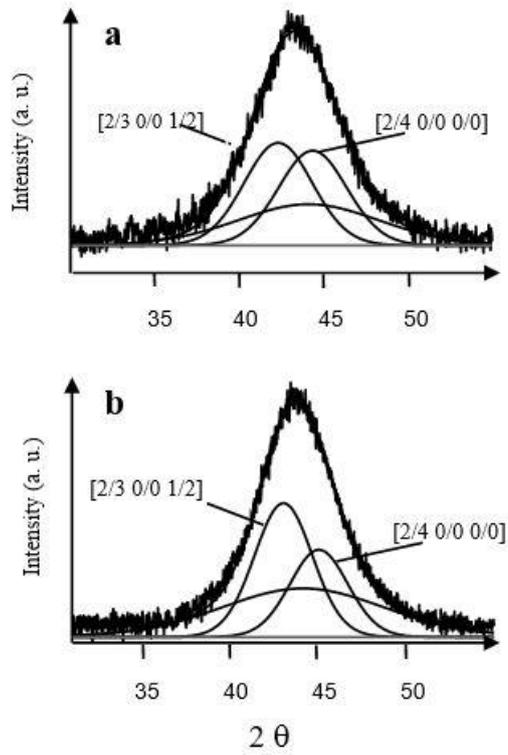

Figure 2 : Fitting of the X-ray diffraction signal obtained on the Ti-Zr-Ni films a/ deposited at 25 °C, b/ deposited at 100 °C. Winfit* software was used (* Winfit 1.2- S. Krumm, Institüt für Geologie-Erlangen). a.u. stands for arbitrary units.

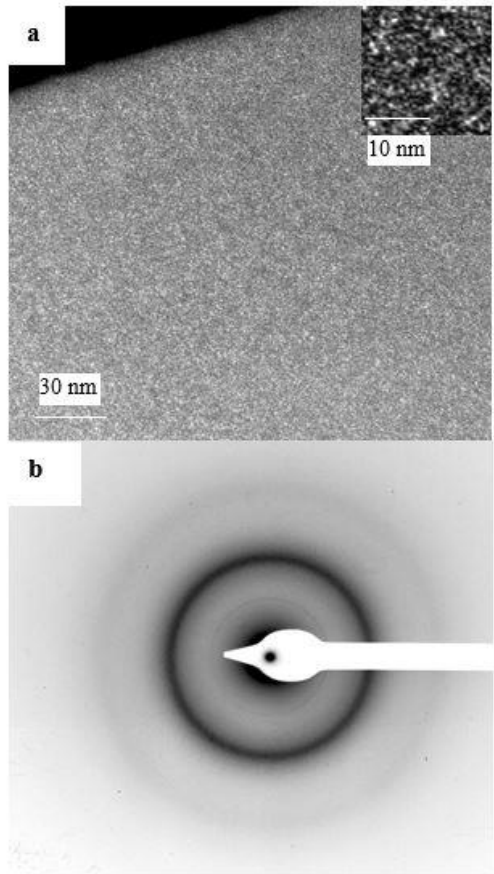

Figure 3: TEM data obtained on a cross-section sample prepared from a film deposited at 100 °C. a/ diffraction pattern. b/ Dark field image evidencing the nano-crystalline nature of the film.

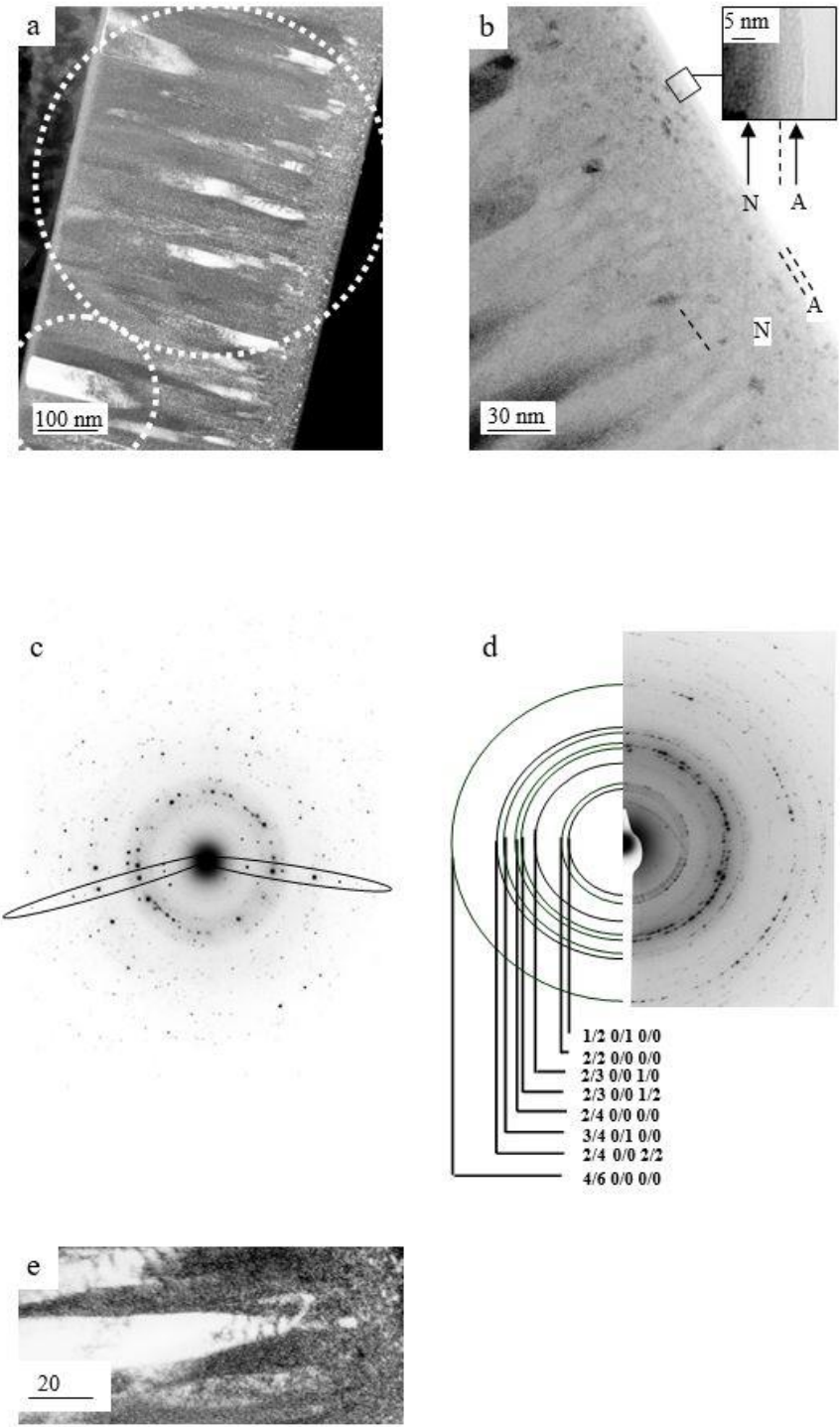

Figure 4: TEM data obtained on a cross-section sample prepared from a film deposited at 275 °C. a/ TEM dark field image using part of the two strongest paraxial rings of diffraction presented in d/ showing a global view of the cross-section, the substrate being located at the bottom right of the image. b/ Bright field image proving the first deposited layers. c/ Diffraction pattern corresponding to the zone laying inside the lower small circle visible in figure a/. d/ Diffraction pattern corresponding to the zone laying inside the upper big circle visible in figure a/.

**Tables**

| Film deposition temperature (°C) | 25 | | | 100 | | |
|---|---|---|---|---|---|---|
| $2\theta$ (°) | 42.290 | 44.330 | 44.085 | 42.973 | 44.988 | 44.085 |
| Indexation | 2/3 0/0 1/2 | 2/4 0/0 0/0 | - | 2/3 0/0 1/2 | 2/4 0/0 0/0 | - |
| FWHM from fitting (°) | 4.8 | 4.8 | 10.0 | 3.9 | 3.9 | 11.0 |
| Real FWHM (°) (experimental widening subtracted) | 4.3 | 4.3 | 9.5 | 3.4 | 3.4 | 10.5 |
| Estimation of the particles size (nm) | ≈ 2 | ≈ 2 | - | ≈ 3 | ≈ 3 | - |

Table 1: Numerical data of peaks obtained by fitting of the X-ray diffraction signal of the 100 °C deposited film pattern. Winfit* software was used (* Winfit 1.2- S. Krumm, Institüt für Geologie-Erlangen).

| Ts | $a_{6D}$ ± 0.001 nm | $a_q$ ± 0.001 nm |
|---|---|---|
| 140 | 0.734 | 0.524 |
| 160 | 0.739 | 0.528 |
| 240 | 0.741 | 0.529 |
| 275 | 0.744 | 0.531 |

Table 2: Ts dependence of the lattice parameter of the icosahedral phase (calculation made on the [2/3 0/0 1/2] peak).